\documentclass[preprint]{elsarticle}
\usepackage{pifont}
\usepackage{natbib}
\usepackage{amssymb}
\usepackage{amsmath}
\usepackage{amsfonts}
\usepackage{amsthm}
\usepackage{mathabx}
\usepackage{MnSymbol}
\usepackage{graphicx}
\usepackage{geometry}

\def\tfract#1/#2{{\textstyle{\raise0.8pt\hbox{$\scriptstyle#1$}\over%
\hbox{\lower0.8pt\hbox{$\scriptstyle#2$}}}}}
\def\mezzo{\tfract 1/2 }

\def\radi2k{\tfract 1/{\sqrt {2k}} }
\def\der{\partial }
\def\downnormalfill{$\,\,\vrule depth4pt width0.4pt
\leaders\vrule depth 0pt height0.4pt\hfill\vrule depth4pt width0.4pt\,\,$}
\def\WT#1{\mathop{\vbox{\ialign{##\crcr\noalign{\kern3pt}
      \downnormalfill\crcr\noalign{\kern0.8pt\nointerlineskip}
      $\hfil\displaystyle{#1}\hfil$\crcr}}}\limits}

\def\be{\begin{equation}}
\def\ee{\end{equation}}
\def\bea{\begin{eqnarray}}
\def\eea{\end{eqnarray}}
\def\ba{\begin{array}{rcl}}
\def\ea{\end{array}}
\def\der{\partial}
\def\go{\leavevmode \raise.3ex\hbox{$\scriptscriptstyle \langle\!\langle\!  $}%
~\ignorespaces}

\def\gf{\relax \ifhmode \unskip~\else \leavevmode \fi \raise.3ex\hbox{$\! \scriptscriptstyle\rangle\!\rangle\, $}}
\begin{document}

\pagestyle{empty}
\setcounter{page}{0}

\begin{center}
{\Large {\bf Topological gauge fixing}}
\\[1cm]
{\large L.~Gallot$^b$, E.~Guadagnini$^a$, E.~Pilon$^b$ and F.~Thuillier$^b$}

\end{center}

\vskip 0.7 truecm

{$^a$ \it Dipartimento di Fisica ``E. Fermi",  Universit\`a di Pisa,  Largo B. Pontecorvo 3, 56127 Pisa; and INFN, Italy.}

{$^b$ \it LAPTh, Universit\'e de Savoie, CNRS, Chemin de Bellevue, BP 110,  F-74941 Annecy-le-Vieux Cedex, France.}

\vspace{1cm}
\begin{center}
{\bf Abstract}
\end{center}

\noindent We implement the metric-independent Fock-Schwinger gauge in the abelian quantum Chern-Simons field theory defined in ${\mathbb R}^3$. The expressions of the various components of the propagator are determined. Although the gauge field propagator differs from the Gauss linking density, we prove that its integral along two oriented knots  is equal to the linking number.

\newpage

The abelian  Chern-Simons  theory in ${\mathbb R}^3$ is a gauge theory defined by the action \cite{1}
\be
S = 2 \pi k \int d^3 x \> \varepsilon^{\mu \nu \rho } A_\mu (x) \partial_\nu A_\rho (x) \; ,
\ee
where $k $ is the real coupling constant of the model and $A_\mu (x)$ denotes the components of the connection. The action (1) is invariant under gauge transformations $A_\mu (x) \rightarrow A_\mu (x) - \partial_\mu \Lambda (x)$. In order to implement the Fock-Schwinger gauge condition \cite{2,3}
\be
x^\mu A_\mu (x) = 0 \; ,
\ee
one can use the BRST procedure in which  the gauge-fixed action $S_{TOT}$,
\be
S_{TOT} = 4 \pi k \int d^3x \, \left [ \mezzo \varepsilon^{\mu \nu \rho} A_\mu (x) \der_\nu A_\rho (x) +  B (x) x^\mu A_\mu (x) + {\overline c} (x) x^\mu \der_\mu c(x) \, \right ] \; ,
\ee
is invariant under the BRST transformations
\be
 \begin{array} {c@{ \quad , \quad }c}
\delta_G \, A_\mu (x) = - \der_\mu c(x) &  \delta_G \, c(x) = 0 \; ,   \\
 \delta_G \, {\overline c}(x) = B(x) &  \delta_G \, B(x) =0 \; .
\end{array}
\ee
The gauge condition (2) does not depend on the metric that one could introduce in ${\mathbb R}^3$, for this reason the Fock-Schwinger condition (2)  is called a topological gauge fixing. Perturbative applications of the gauge (2) in ordinary gauge theories of the Yang-Mills type
have been discussed in Ref.\cite{4,5,6}, whereas applications in the description of nonperturbative effects can be found in \cite{7}.

A nice feature of the gauge (2) is that
$A_\mu (x)$ can be expressed in terms of the curvature $F_{\mu \nu} = \der_\mu A_\nu - \der_\nu A_\mu $  by means of the homotopic  formula
\be
A_\mu (x) =   \int_0^1 d s  \, s   x^\nu F_{\nu \mu }(s x ) \; .
\ee
Consequently the propagator of the field $A_\mu (x)$ can directly be derived  from the two-point function $ \langle F_{\mu \nu } (x)  F_{\rho \lambda } (y) \rangle$ of the curvature.
Since $F_{\mu \nu }$ is gauge-invariant,  the expectation value $ \langle F_{\mu \nu } (x)  F_{\rho \lambda } (y) \rangle$ does not depend on the choice of the gauge and is determined  by the Chern-Simons action (1) exclusively. In order to compute the two-points function of the curvature, there is no need to fix the gauge;  indeed the variation of  the action (1) gives
\be
\varepsilon^{\mu \nu \rho} \der_\nu A_\rho (x)= (1/ 4 \pi k )\, \delta S[A] / \delta A_\mu (x) \; .
\ee
Therefore one obtains
\bea
\langle \varepsilon^{\mu \nu \rho} \der_\nu A_\rho (x) \varepsilon^{\lambda \sigma \tau } \der_\sigma A_\tau (y) \rangle &=& {\int DA \, e^{iS[A]} \varepsilon^{\mu \nu \rho} \der_\nu A_\rho (x) \varepsilon^{\lambda \sigma \tau } \der_\sigma A_\tau (y)  \over \int DA \, e^{iS[A]} } \nonumber \\
&=& - \left (i \over 4 \pi k \right ) {\int DA  \left ( \delta e^{iS[A]} / \delta A_\mu (x) \right )  \varepsilon^{\lambda \sigma \tau } \der_\sigma A_\tau (y)  \over \int DA \, e^{iS[A]} } \nonumber \\
&=& \left (i \over 4 \pi k \right ) {\int DA  \, e^{iS[A]}   \varepsilon^{\lambda \sigma \tau } \left ( \delta [ \der_\sigma A_\tau (y)] / \delta A_\mu(x) \right ) \over \int DA \, e^{iS[A]} } \nonumber \\
&=&   \left ( {i \over 4\pi k} \right ) \, \varepsilon^{\lambda  \sigma \mu} \, { \der \over \der y^\sigma } \delta^3 (x-y) \; .
\eea
Being gauge independent, the result (7) can also be checked in any particular gauge, like for instance the covariant Landau gauge \cite{8}.
Equations (5) and (7)  determine the expression of the Feynman propagator for the connection $A_\mu (x)$ in the Fock-Schwinger gauge
\bea
\WT{A_\mu (x)\> A}\! \null_\nu (y)  &=& \int_0^1 \! dt \int_0^1 \! ds \, t x^\lambda \, s y^\tau \, \langle F_{\lambda \mu} (t x) F_{\tau \nu} (s y) \rangle \nonumber \\ &=&
\left ( {-i \over 4 \pi k } \right ) \varepsilon_{\lambda \mu \alpha } \varepsilon_{ \tau \nu \beta } \varepsilon^{\alpha \beta \rho }\int_0^1 \! dt \int_0^1 \! ds \, t x^\lambda \, s y^\tau   {\der \over \der z^\rho}  \delta^3 (z - s y)  \bigg |_{z= t x} \; .
\eea
The propagator (8) is symmetric under the combined exchanges  $x \leftrightarrow y $ and $\mu \leftrightarrow \nu $.

As in any abelian gauge theory, the anticommuting ghost and antighost fields $c(x)$ and ${\overline c} (x)$ decouple from the remaining fields and can be ignored. Similarly, the auxiliary field  $B(x)$ does not enter into the interaction lagrangian; so $B(x)$ also can be ignored. Nevertheless,   let us derive for completeness the remaining components of the propagator, which can also be used in the case of an  interacting theory. The gauge-fixed action (3) determines the five equations which must be satisfied by the various components of the propagator in the Fock-Schwinger gauge:
\be
\varepsilon^{\mu \nu \rho } {\der \over \der x^\nu } \WT{A_\rho (x)\> A}\! \null_\lambda (y) + x^\mu \WT{B (x)\> A}\! \null_\lambda (y) = \left ( {i \over 4 \pi k } \right ) \delta^\mu_\lambda \, \delta^3 (x-y) \; ,
\ee
\be
\varepsilon^{\mu \nu \rho } {\der \over \der x^\nu } \WT{A_\rho (x)\> B}  (y) + x^\mu \WT{B (x)\> B} (y) = 0 \; ,
\ee
\be
x^\rho \WT{A_\rho (x)\> A}\! \null_\lambda (y) = 0 \; ,
\ee
\be
x^\rho \WT{A_\rho (x)\> B}  (y) = \left ( {i \over 4 \pi k } \right ) \delta^3 (x-y) \; ,
\ee
\be
x^\mu {\der \over \der x^\mu } \WT{c (x)\> {\overline c}}  (y) = \left ( {i \over 4 \pi k } \right ) \delta^3 (x-y) \; .
\ee
Expression (8) fulfils  equation (11);  equations (9), (10) and (12) confirm the validity of equation (8) and fix the remaining components of the propagator for the fields of commuting type
\bea
&& \WT{B (x)\> A}\! \null_\lambda (y)  = \left ( {i \over 4 \pi k } \right )  \int_0^1 d t \,  t {\der \over \der x^\lambda } \delta^3 (t x - y)  \; , \nonumber \\  && \WT{B (x)\> B} (y) = 0 \; .
\eea
Finally equation (13) specifies the ghost-antighost propagator
\be
\WT{c (x)\> {\overline c}}  (y) =  \left ( {- i \over 4 \pi k } \right ) \int_0^1 \! dt \, t^2 \, \delta^3 (x- ty) \; .
\ee

The expectation values of the gauge holonomies associated with knots are the basic observables of the Chern-Simons theory; since each  holonomy computed along a closed path is gauge invariant, these observables do not depend on the choice of the gauge. In particular,  it is known \cite{8,9,10} that in the abelian CS theory the integral of the $A_\mu $ propagator along two oriented knots $C_1$ and $C_2$ in ${\mathbb R}^3$ gives the corresponding linking number, according to  the following  relation
\be
\oint_{C_1} dx^\mu \oint_{C_2} dy^\nu \, \WT{A_\mu (x)\> A}\! \null_\nu (y)  = \left ( { i \over 4 \pi k } \right ) \, \ell k (C_1 , C_2) \; .
\ee
Equations (8) and (16) imply
\be
\ell k (C_1 , C_2) = \mezzo \int_0^1 \! d t  \oint_{C_1} \! dx^\mu \oint_{C_2} \! d y^ \nu \,  \varepsilon_{\mu \nu \lambda } \,  \left [ y^\lambda \delta^3 (t x - y) -  x^\lambda \delta^3 (x - t y) \right ] \; .
\ee
Let us verify that formula (17) gives a representation of the linking number. The function that has to be integrated in expression  (17) does not vanish only when $t x^\mu = y^\mu $  or $t y^\mu = x^\mu $. These two possibilities are mutually exclusive alternatives  because $t $ varies in the range $0 \leq t \leq 1$. In general, equations $t x^\mu = y^\mu $  or $t y^\mu = x^\mu $ are satisfied in a finite number of isolated points in the variables space; if this is not the case, then one can slightly  modify the knots $C_1$ and $C_2$ by an ambient isotopy  in such a way to  put them  in a generic position.

Let us introduce a parametrization $x^\mu = x^\mu (\xi ) $ (with $-1 \leq \xi \leq 1$) for the knot $C_1$ and a parametrization  $y^\nu = y^\nu (\zeta )$ (with $-1 \leq \zeta \leq 1$) for the knot  $C_2$. Suppose that when $\xi =0$ and $\zeta =0 $ one finds   for instance $t_0 x^\mu(0) = y^\mu(0)$  for a given  $ 0 < t_0 < 1$. Without loss of generality, one can choose a cartesian reference system so that, in a neighborhood of $\xi =0$ and $\zeta =0$ in the variables space, one has
\be
 \left \{  \begin{array} {r@{ = }l}
 x^1 (\xi ) &  \xi a + {\cal O}(\xi^2)   \\
x^2(\xi ) &  0 + {\cal O}(\xi^2)   \\
x^3 (\xi ) & b + {\cal O}(\xi^2)
\end{array}  \right.  \quad , \quad
\left \{  \begin{array} {r@{ = }l}
 y^1 (\zeta ) &    \zeta c + {\cal O}(\zeta^2) \\
y^2(\zeta ) &  \zeta h + {\cal O}(\zeta^2) \\
y^3 (\zeta ) &    t_0 b + \zeta g + {\cal O}(\zeta^2)
\end{array} \right.
\ee
for certain real constants $a $, $b$, $c$, $h$ and $g$. The components  $\dot x^\mu ( \xi ) = d x^\mu (\xi ) / d \xi $ and $\dot y^\nu (\zeta ) = d y^\nu (\zeta ) / d \zeta $ are given by
\be
\left \{  \begin{array} {r@{ = }l}
\dot x^1 (\xi ) & a + {\cal O}(\xi )   \\
\dot x^2(\xi ) &  0 + {\cal O}(\xi ) \\
\dot x^3 (\xi ) & 0 + {\cal O}(\xi )
\end{array} \right. \quad , \quad
\left \{  \begin{array} {r@{ = }l}
 \dot y^1 (\zeta ) &  c  + {\cal O}(\zeta ) \\
 \dot y^2(\zeta ) & h   + {\cal O}(\zeta ) \\
 \dot y^3 (\zeta ) &  g + {\cal O}(\zeta )
 \end{array} \right.
 \ee
 Since the knots $C_1$ and $C_2$ do not intersect, then $b \not= 0$. Moreover,  with good  parametrizations  $x^\mu ( \xi )$ and $y^\nu (\zeta )$ of the knots,  $a \not=0$ and $(c^2 + h^2 + g^2)^{1/2} \not= 0$. Possibly with the introduction of a small deformation on $C_2$, one also ensures that $h \not= 0$. Let $\mezzo w(0)$ be the contribution to the expression (17) coming from the integration in a  neighborhood  of $\xi =0 $ and $\zeta =0 $; equations (18) and (19) determine the local expression of  $\mezzo w(0)$,
\bea
 \mezzo w (0) = \mezzo \int \! && {\hskip -0.63 truecm} d t \! \int \! d\xi  \! \int \! d \zeta   \, \varepsilon_{1 2 3 } \,  ( a + \cdots ) (  h + \cdots )   ( t_0 b   + \cdots )  \times \nonumber \\
 &&   \times \delta (t \xi  a  - \zeta c + \cdots ) \delta ( \zeta h  + \cdots )  \delta (t b - t_0 b  + \cdots )  \; .
\eea
The integral  in $d t $ is fixed by the delta function $\delta (t b - t_0 b  + \cdots ) = \delta (t - t_0 ) / | b| $, and the integral in $d\zeta $ is saturated by the delta function $\delta (\zeta h  + \cdots ) = \delta (\zeta ) / |h |  $. Finally the integral in $d \xi $ eliminates the last delta function $ \delta (t \xi  a  - \zeta c + \cdots ) = \delta (t_0 \xi a   + \cdots ) = \delta ( \xi ) /  t_0  \, | a |   $,
thus
\be
 w (0) =  \varepsilon_{1 2 3 }  \, { a \over |a| } \, {h \over |h| }\, { b \over |  b | }  = \pm  1 \; .
\ee
The sign $ a / |a| $ determines  the direction of $\dot x^1 (0)$ (with respect to the oriented $\widehat 1$-axis of a cartesian system) and $h / |h| $ specifies the direction of $\dot y^2 (0)$ (with respect to the oriented $\widehat 2$-axis of a cartesian system). The ratio $b / |b| $ determines the  position of $x^3(0)$ in the $\widehat 3$-axis with respect to the origin of the reference system; this is equivalent to fix the orientation of the $\widehat 3$-axis with respect to a right-handed system in which $\varepsilon_{1 2 3} = + 1$.

Let us consider  a right-handed cartesian reference system; if the knots $C_1$ and $C_2$ are represented by a two-dimensional diagram which is obtained by a projection of $C_1 \cup C_2 \subset {\mathbb R}^3$ on the $ (\, \widehat 1 , \widehat 2 \, )$-plane, the two possible values of $w(0)$ are shown in Figure~1.

\vskip 1.5 truecm
\centerline {\includegraphics[width=3.01 in]{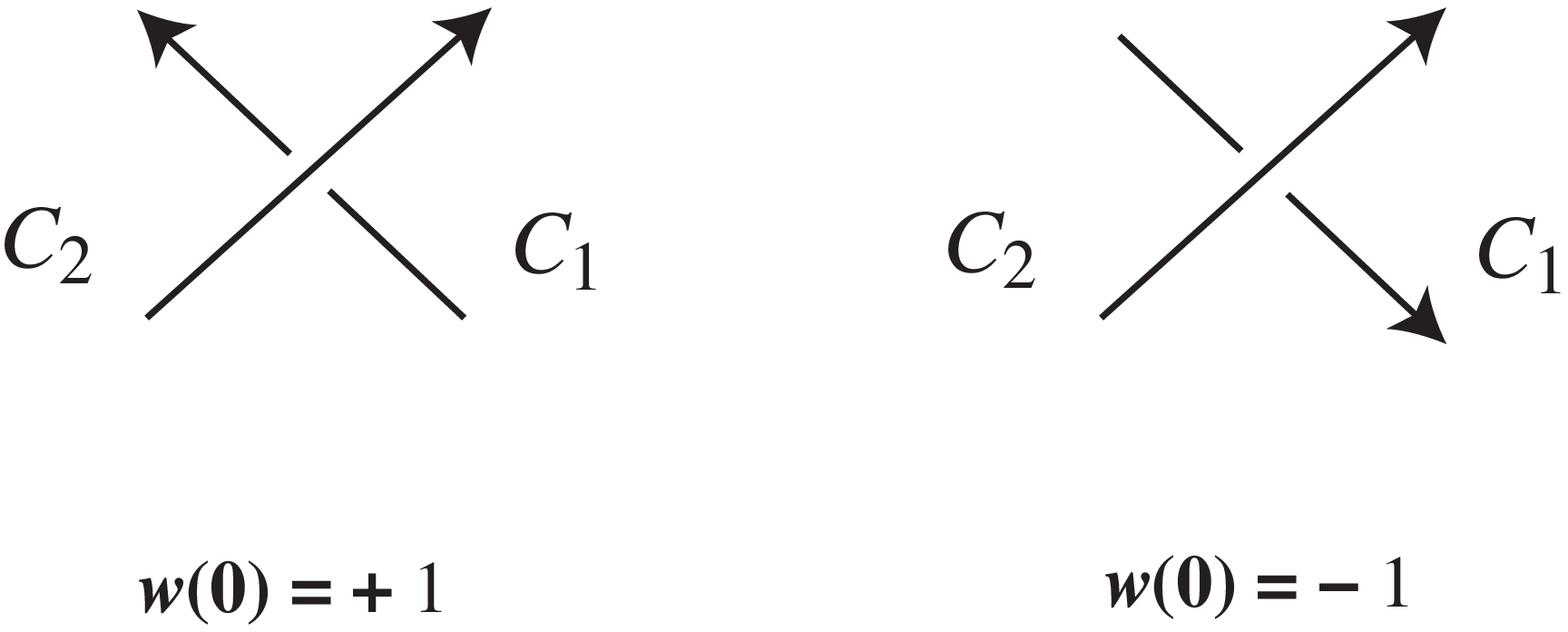}}
\vskip 0.4 truecm
\centerline {{Figure 1.} {Value of $w(0)$ at the  crossing point in a link diagram.}}

\vskip 0.6 truecm

\noindent The crossing point in the diagram corresponds to the vanishing values of the parameters, $\xi =0$ and $ \zeta =0$, for which $ t_0 x^\mu (0) = y^\mu (0)$. The overall convention on the sign of $w(0)$ ---with respect to the crossing configuration in a link diagram---  is fixed by the assumption that the origin of the right-handed cartesian system is placed above the plane of the diagram (and coincides with one eye of the reader).  In the case in which $ t_0 y^\mu (0) = x^\mu (0)$, the values of $w(0)$ are determined again by Figure~1 in which one has simply to exchange $C_1$ with $C_2$.

 Then the whole integral (17) is given by the sum of the contributions of type (21) coming from  all the points in the variables space for which $ t x = y $ or $t y = x$. That is, given a link diagram $D_L$ that represents the link $L = C_1 \cup C_2$,  the value of the whole integral (17) is equal to  the sum of the numbers $\left  \{ \mezzo w(p) \right \} $ over all the crossing points $\{ p \}$ between the two subdiagrams $D_1$ and $D_2$ ---which describe $C_1$ and $C_2$ respectively--- of the link diagram $D_L$
\be
 \mezzo \int_0^1 \! d t \oint_{C_1} \! dx^\mu \oint_{C_2} \! d y^ \nu \,  \varepsilon_{\mu \nu \lambda } \,  \left [ y^\lambda \delta^3 (t x - y) -  x^\lambda \delta^3 (x - t y) \right ] = \mezzo \sum_{p\in D_1 \cap D_2} w (p) \; .
\ee
Since the writhe number [11] of a link diagram $D_L$ is given by $w(D_L) = \sum_{p \in D_L} w(p)$,  expression (22) can be written  as the difference of the writhe numbers $w(D_L)$ and $w(D_1) + w(D_2)$,
\be
\mezzo \sum_{p\in D_1 \cap D_2} w (p) = \mezzo \left [ w(D_L) - w (D_1) - w(D_2) \right ]  \; ,
\ee
and it is known \cite{11} that  expression (23) gives precisely the linking number of $C_1 $ and $C_2$. This concludes the proof of the validity of equation (17).

The Fock-Schwinger gauge (2) also admits an abstract formulation which is defined in terms of a suitable homotopy operator $h$ acting on differential forms. The corresponding formal construction can easily be generalised in higher dimensions. Let us sketch out this idea, the details will be discussed in a forthcoming article.

In the space $\Omega^p$  of $p$-forms in ${\mathbb R}^n$,  the Poincar\'e homotopy operator $h \, : \Omega^p \rightarrow \Omega^{p-1}$ (with $p > 0$) is defined by \cite{12}
\be
(h \omega)(x)
= {1 \over (p-1)!} \left( \int_{0}^{1} dt \, t^{p-1} \, x^{\nu} \, \omega_{\nu \mu_2 \cdots \mu_p}(tx) \right) \, dx^{\mu_2} \wedge \cdots \wedge dx^{\mu_p} \; ,
\ee
and satisfies $h^2 =0 $ together with the fundamental identity
\be
d h + h d = 1 \; .
\ee
The Poincar\'e homotopy gauge on $A \in \Omega^1$ is defined by the condition
\be
h A = 0 \; ,
\ee
which is the analogue of equation (2). Let ${\cal A}_h$ be space of gauge potentials satisfying equation (26); any element $A$ of $ {\cal A}_h$ can be written as
\be
A = (d h + h d ) A = h d A = h F \; ,
\ee
which coincides with equation (5). Note that,  since $d F = d d A = 0$, one has
\be
d A = F = (d h + h d ) F = d h F \; ,
\ee
which is consistent with equation (27). The correspondence (27) can be used to express the Chern-Simons action (1)  in terms of the curvature $F = dA$,
\be
 S = 2 \pi k \int h F \wedge F = \left ( 4 \pi k \right ) \int \mezzo \left ( h F \wedge  F \right ) \; .
\ee
Since the $A$ and $F$ fields are linearly related, expression (29)  determines the propagator for the field $F$ according to the standard rule
\be
\langle F(x) F(y) \rangle \sim \left ( {i \over 4 \pi k} \right ) \, h_y^{-1} \delta^3 (x-y) \; .
\ee
Equation (25) shows that, in the space of closed forms, the $d$ operator is the inverse of the homotopy operator; this implies
\be
\langle F(x) F(y) \rangle \sim \left ( {i \over 4 \pi k} \right ) \, d_y \, \delta^3 (x-y) \; ,
\ee
which is really equivalent of equation (7). Finally  one gets
\be
 \langle A(x) A(y) \rangle = \langle hF(x) \, h F(y) \rangle \sim \left ( {i \over 4 \pi k} \right ) \, h_x \delta^3 (x-y) \; ,
\ee
that, when field indices are written explicitly,  takes the form shown in equation (8).

\vskip 1.4 truecm

\bibliographystyle{amsalpha}

\end{document}